\documentclass[conference]{IEEEtran}
\IEEEoverridecommandlockouts
\usepackage{cite}
\usepackage{amsmath,amssymb,amsfonts}
\usepackage{graphicx}
\usepackage{textcomp}
\usepackage{xcolor}
\usepackage{mathtools}
\usepackage[linesnumbered,algoruled]{algorithm2e}
\usepackage{algpseudocode}
\usepackage{amsthm}
\usepackage{comment}
\usepackage{bm}
\usepackage{chemformula}
\usepackage{tabularx}
\usepackage{graphicx}
\usepackage{mwe}
\usepackage{subfig}
\usepackage{float}
\usepackage{siunitx}
\usepackage{xparse}
\DeclareSIUnit{\pH}{pH}
\DeclareSIUnit \voltampere { VA } 
\DeclareSIUnit \var { var } 
\DeclareSIUnit\mt{\milli\tesla} 
\usepackage[all]{background}

\usepackage{stackengine}
\setstackEOL{\\}
\setstackgap{L}{\normalbaselineskip}
\SetBgContents{\color{gray}{\tiny \Longstack{PREPRINT - accepted by IEEE International Symposium on Hardware Oriented Security and Trust (HOST) 2022}}}
\SetBgPosition{3.5cm,1cm}
\SetBgOpacity{1.0}
\SetBgAngle{0}
\SetBgScale{1.8}

\def\BibTeX{{\rm B\kern-.05em{\sc i\kern-.025em b}\kern-.08em
    T\kern-.1667em\lower.7ex\hbox{E}\kern-.125emX}}
\begin{document}
\bstctlcite{IEEEexample:BSTcontrol}
\title{pHGen: A pH-Based Key Generation Mechanism Using ISFETs \vspace{-2.5mm}}
\author{\IEEEauthorblockN{Elmira Moussavi\IEEEauthorrefmark{1},
Dominik Sisejkovic\IEEEauthorrefmark{1},
Fabian Brings\IEEEauthorrefmark{2},
Daniyar Kizatov\IEEEauthorrefmark{1},
Animesh Singh\IEEEauthorrefmark{2},\\
Xuan Thang Vu\IEEEauthorrefmark{2},
Rainer Leupers\IEEEauthorrefmark{1}, 
Sven Ingebrandt\IEEEauthorrefmark{2},
Vivek Pachauri\IEEEauthorrefmark{2}, and
Farhad Merchant\IEEEauthorrefmark{1}}
		\IEEEauthorblockA{ \IEEEauthorrefmark{1}Institute for Communication Technologies and Embedded Systems, \IEEEauthorrefmark{2}Institute of Materials in Electrical Engineering 1}
        \IEEEauthorblockA{RWTH Aachen University, Germany}
        \{moussavi, sisejkovic, kizatov, leupers, merchantf\}@ice.rwth-aachen.de \\ \{brings, singh, vu, ingebrandt, pachauri\}@iwe1.rwth-aachen.de \vspace{-5mm} }
\maketitle

\begin{abstract}
Digital keys are a fundamental component of many hardware- and software-based security mechanisms. However, digital keys are limited to binary values and easily exploitable when stored in standard memories. In this paper, based on emerging technologies, we introduce pHGen, a potential-of-hydrogen (pH)-based key generation mechanism that leverages chemical reactions in the form of a potential change in ion-sensitive field-effect transistors (ISFETs). The threshold voltage of ISFETs is manipulated corresponding to a known pH buffer solution (key) in which the transistors are immersed. To read the chemical information effectively via ISFETs, we designed a readout circuit for stable operation and detection of voltage thresholds. To demonstrate the applicability of the proposed key generation, we utilize pHGen for logic locking---a hardware integrity protection scheme. The proposed key-generation method breaks the limits of binary values and provides the first steps toward the utilization of multi-valued voltage thresholds of ISFETs controlled by chemical information. The pHGen approach is expected to be a turning point for using more sophisticated bio-based analog keys for securing next-generation electronics.

\end{abstract}

\begin{IEEEkeywords}
	Hardware security, emerging technologies, logic locking, key generation, ion-sensitive field-effect transistor.
\end{IEEEkeywords}

\section{Introduction}
Security has recently become particularly important for
hardware designs~\cite{5604160}. Especially, due to the high cost of manufacturing and maintaining nanoscale integrated circuit (IC) foundries, a significant portion of chips are designed and fabricated offshore~\cite{ravi2004security}. This globalization of the IC design flow often leaks the vital information of the chips to untrustworthy parties and creates opportunities for intellectual property (IP) theft, overproduction, reverse engineering, counterfeiting, and insertion of hardware Trojans~\cite{rostami2014primer}. Techniques to ensure the security of hardware integrity through IC design and fabrication have been one of the emerging areas of information technology in recent years, where hardware security approaches utilize conventional digital keys to offer protection from an attack. However, there remain several ways to attack, such as exploitation of the digital data (e.g., secret keys in cryptographic applications) enabled by the highly vulnerable storage.

\begin{figure}[!t]
   \centering
   \includegraphics*[width=\columnwidth]{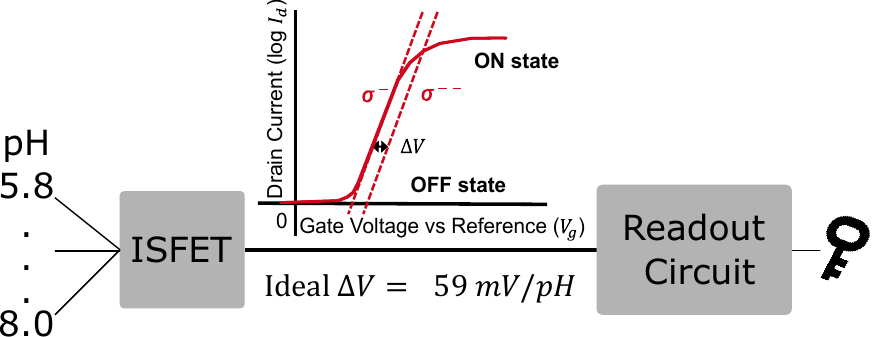}
   \caption{The flow of pHGen.}
   \label{fig:pHGen}
   \vspace{-5mm}
\end{figure}

In this work, we present pHGen, a pH-based key generation mechanism that utilizes emerging transistor technologies to generate secure keys from (bio)chemical information. The bio-based key generation offers multiple advantages over standard, binary keys, including a multi-valued representation, enhanced security in terms of physical key inspection, and a path towards personalized, biochemical digital keys. The main contributions of this paper are as follows (Fig.~\ref{fig:pHGen}):
\begin{itemize}
    \item	The implementation of pHGen; a pH-based key generation mechanism that manipulates the voltage thresholds of ISFETs immersed into specific pH buffer solutions to extract analog chemical keys.
    \item The design and development of a stable readout circuit to read the intrinsic voltage threshold shift of ISFETs for integration with field-programmable gate arrays (FPGAs).
    \item The demonstration of pHGen for the generation of secure analog keys in logic-locked circuits.

\end{itemize}

\section{pHGen Implementation}
\subsection{ISFET- an Emerging Technology}
To perform the transition from digital to analog keys, we utilize hydrogen ion (H+) concentration or pH values as they reflect chemical information. Miniaturized ion-selective sensors such as micro and nanoISFETs can be used as rapid analytical devices to detect small variations of free hydrogen and hydroxyl ions\cite{CAENN201755}. In this work, we use microISFETs as a test-board for well-established, reproducible, clean-room fabrication at wafer-scale, ion-recognition, and circuit-integration possibilities. Sensitive to surface-charge, ISFETs are utilized as chemical and biosensors to analyze solution samples in contact with its gate \cite{6809216}. In addition, ISFETs offer possibilities for scalable integration using an unmodified CMOS process towards industrial application. ISFETs operate similar to MOSFETs, whose gate is extended with a passivation layer in contact with an external reference electrode \ch{(Ag/AgCl)} via a sample solution (Fig. \ref{fig:ISFET}) also known as electrochemical gate configurations. The output current $I_{DS}$ of an n-type transistor is expressed as follows:
\begin{figure}[!t]
   \centering
   \includegraphics*[width=0.91\columnwidth]{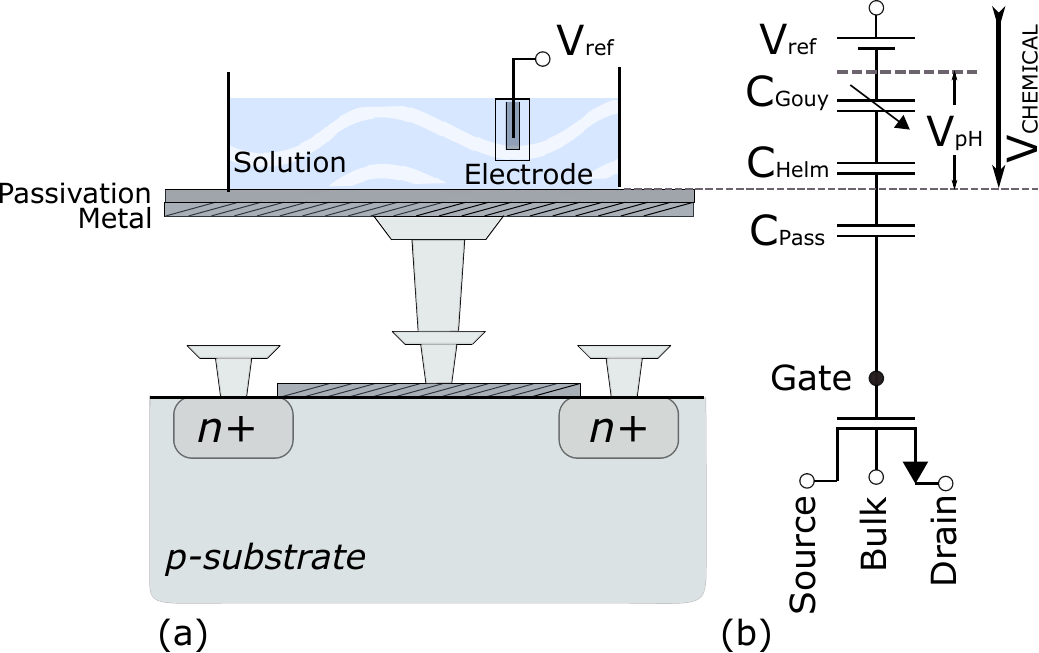}
   \caption{(a) ISFET as a modified MOSFET technology and (b) equivalent electrical model.}
   \label{fig:ISFET}
   \vspace{-5mm}
\end{figure}
\begin{equation}
    I_{DS}=\mu_{n}c_{ox}\frac{W}{L}(V_{GS}-V_{TH(ISFET)})V_{DS}-\frac{1}{2}V_{DS}^{2}.
    \label{eq1}
\end{equation}

In comparison to MOSFET, $V_{TH(MOSFET)}$ is replaced by $V_{TH(ISFET)}$ in eq. \ref{eq1}, which is influenced by surface charge changes as a function of the pH value of the solution in contact with the ISFET gate. Thus, ISFETs measure the solution pH in the form of threshold voltage change ($V_{TH}$) of a corresponding floating-gate MOSFET device. The electrochemical interface ($V_{CHEMICAL}$) consists of a reference built-in potential ($V_{ref}$) together with the double-layer capacitance known as Gouy–Chapman ($C_{Gouy}$) and Helmholtz ($C_{Helm}$) plane\cite{van1995novel}. $C_{Gouy}$ and $C_{Helm}$ define the pH potential drop ($V_{pH}$) at the MOSFET gate and the $C_{pass}$ is the passivation capacitance. The electrochemically induced voltage in the $V_{TH(ISFET)}$ can be defined as follows:
\begin{equation}
    V_{TH(ISFET)}=V_{TH(MOSFET)}-V_{CHEMICAL},
\end{equation}
where $V_{CHEMICAL}$ is defined as:
\begin{equation}
    V_{CHEMICAL}=E_{i}+\frac{RT}{n_{i}F}ln(\alpha_{i}).
    \label{eq3}
\end{equation}
In eq. \ref{eq3}, $E_{i}$ is the standard electrode potential, R: gas constant , T: absolute temperature (K), F: Faraday constant, $n_{i}$: charge of ion $i$, and $\alpha_{i}$: activity of ions. The only variable in eq. \ref{eq3} is $\alpha_{i}$. Thereby, to monitor the ion activity it is sufficient to keep $V_{DS}$ and $I_{DS}$ constant. In this case, $V_{GS}$ is proportional to the natural logarithm of ion activity. Hence, the ISFET needs to be connected to a readout circuit to provide this condition.

\subsection{Readout Circuit Design}
\begin{figure}[!t]
   \centering
   \includegraphics*[width=1\columnwidth]{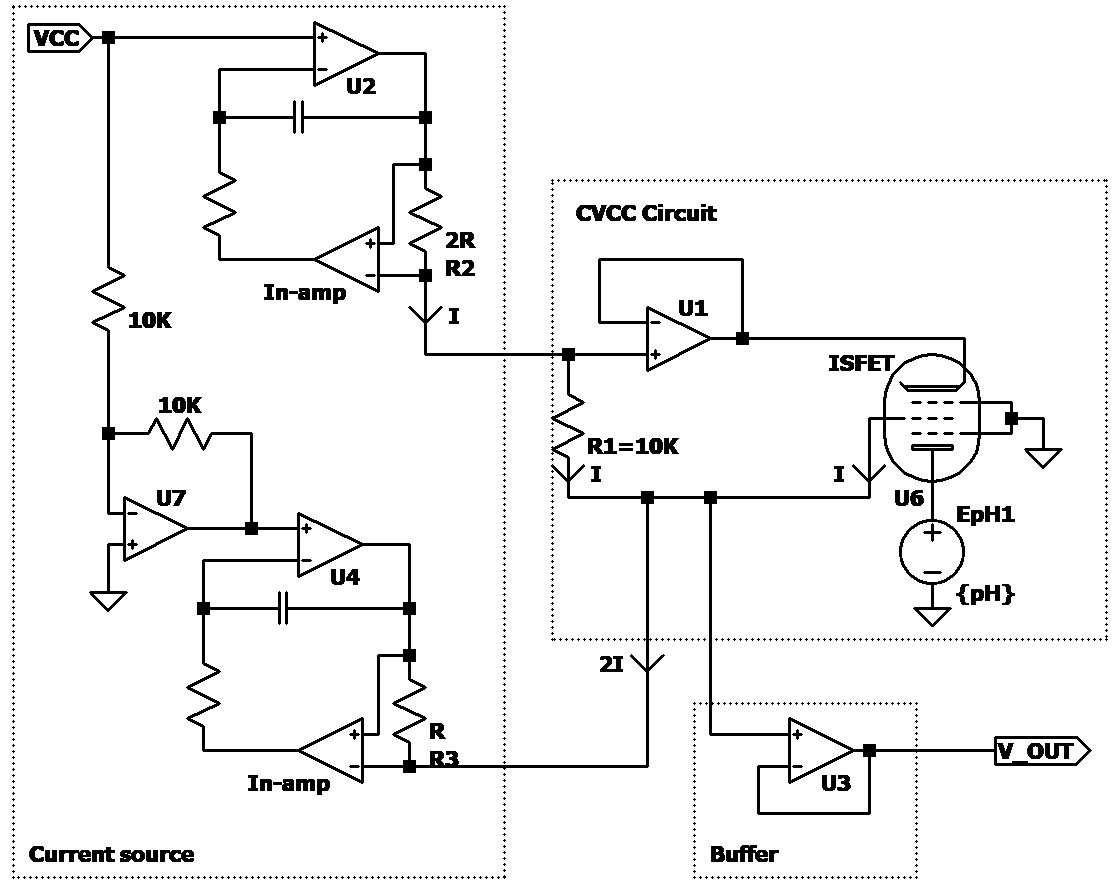}
   \caption{Readout circuit designed for ISFET integration.}
   \label{fig:isfetroc}
   \vspace{-5mm}
\end{figure}

A constant voltage constant current (CVCC) circuit is implemented as a readout circuit (ROC) to monitor the ion activity in response to an electrolyte solution that is in contact with the ISFET gate. For robustness, the sensor operates in the linear region, and the gate voltage is connected to the ground. Thus, the source voltage serves as the output signal proportional to the internal ISFET’s threshold voltage ($V_{TH(ISFET}$). Fig. \ref{fig:isfetroc} shows the designed ROC. The voltage across ISFET's drain and source ($V_{DS}$) is constant and equal to $R1*I$. In addition, the source current is half to that of the sink current, which according to Kirchhoff's current law, keeps $ I_ {DS} $ constant and is equivalent to $ I $. A current source circuit is used to generate the current sources. This circuit includes an instrumentation amplifier (In-amp) and an operational amplifier. Therefore, the CVCC circuit ensures a constant operating current over ISFET's drain-source terminals ($I_{DS}= I =$ \SI{50}{\micro\ampere}). Consequently, R1 constantly maintains $V_{DS} = R1*I =$ \SI{10}{\kilo\ohm} $*$ \SI{50}{\micro\ampere} $=$ \SI{0.5}{\volt}. In addition to current source circuits and the R1 resistance, the CVCC circuit comprises one operational amplifier (U1) connected to ISFET's drain providing feedback to the source and ensuring the functionality of the CVCC circuit. A buffer stage is employed to improve driving capability. Drawbacks related to temperature dependency~\cite{gui1987isfet} can be addressed by using a temperature sensor to compensate for the temperature effect.

\section{Demonstration}
\begin{figure*}[!t]
\vspace{-5mm}
\centering
   \subfloat[ISFET pH sensitivity with \ch{(Ta2O5)} gate at the temperature of 30\textdegree{}C.]{\includegraphics[width=1\columnwidth]{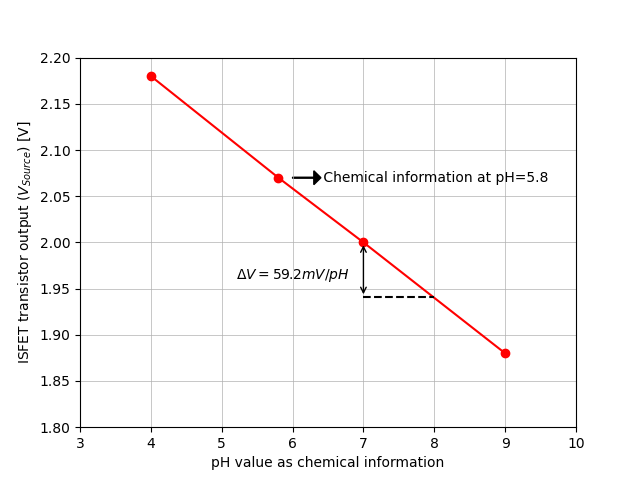}}
    \subfloat[ Reading the pH response and temperature information from the (bio)chemical-system.]{\includegraphics[width=1\columnwidth]{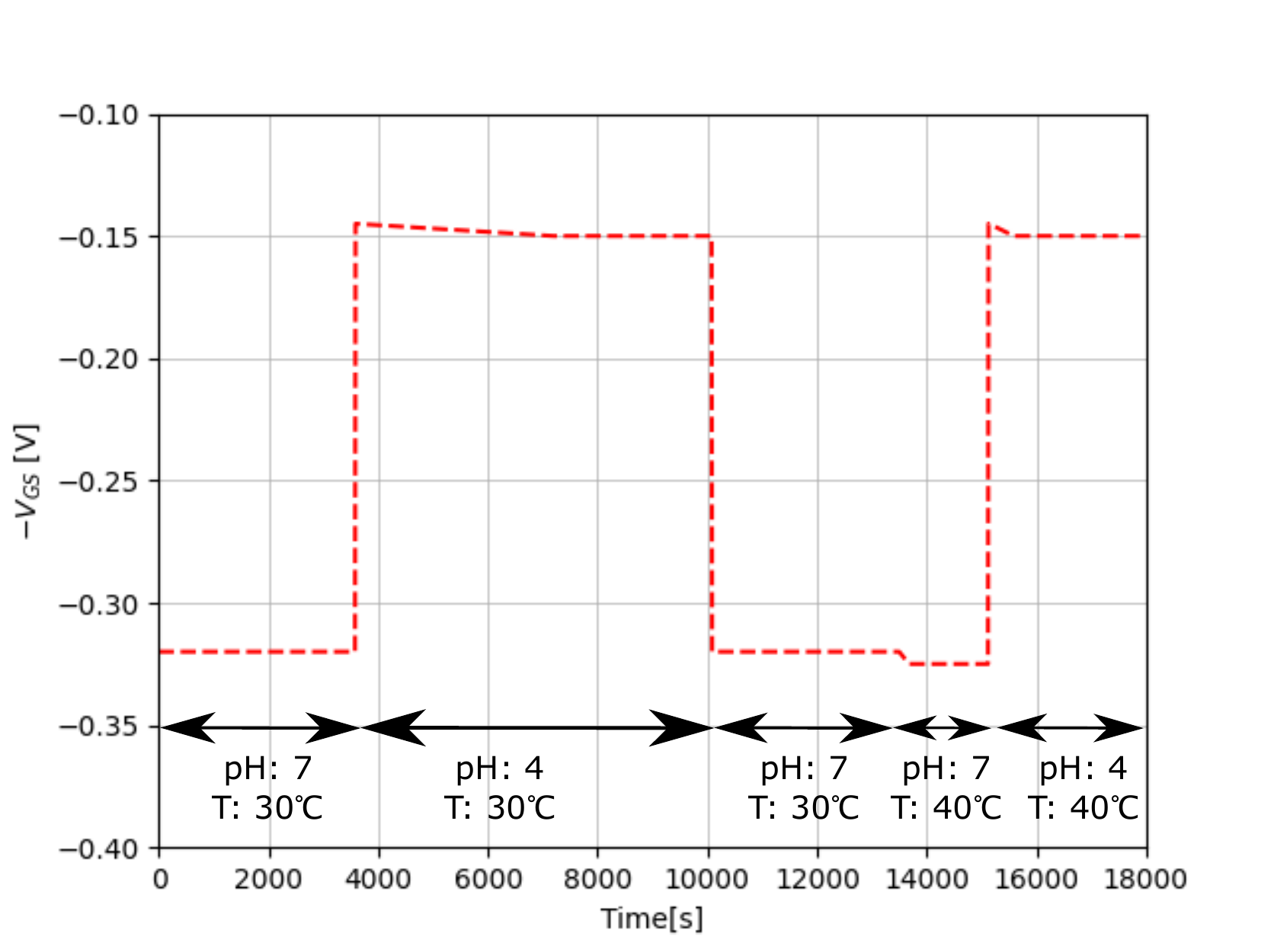}}
    \caption{Results for (a) source voltage changes vs. pH solutions, and (b) gate source voltage change over the time at constant drain source voltage of 0.5V. }\label{fig:foobar}
    \vspace{-5mm}
\end{figure*}
\begin{table}[!b]
\caption{\label{tab:isfet} MSFET 3330-2 pH sensor dimensions.}
\centering
\begin{tabular}{c|c|c|c}
\hline
                 & Width & Height & Length \\ \hline
Chip dimensions &  \SI{1.2}{\milli\meter} & \SI{0.3}{\milli\meter} & \SI{3}{\milli\meter}    \\ \hline
Packaged sensor  & \SI{5}{\milli\meter}   &  \SI{1}{\milli\meter} & \SI{50}{\milli\meter}   \\ \hline
\end{tabular}
\end{table}
Key-driven countermeasures, such as logic locking, have been proposed to protect the integrity of hardware designs through the IC design and fabrication flow~\cite{yasin2020trustworthy, scopes2020, date2021}. Logic locking inserts a locking mechanism that ensures a correct circuit behavior only upon the application of a correct activation key. These inserted key-dependent gates are referred to as key gates.
As the key is loaded post-fabrication by the IP owner, the design remains concealed while in the hands of untrusted third parties such as the external design house and foundry.
To demonstrate the applicability of pHGen, we present a logic-locked circuit that uses the pH solution as the source of the input key to drive an XOR key gate. The demonstration was implemented with an MSFET 3330-2 ISFET, which is a commercial ultra-miniature pH ISFET sensor from MICROSENS S.A~\cite{Microsens}. It consists of silicon- and polysilicon-based materials and a \ch{Ta2O5} gate as a pH-sensitive metal with 4\ensuremath{^{\prime\prime}} planar CMOS process technology. The ISFET sensor dimensions are listed in Table \ref{tab:isfet}. The voltage threshold shift per unit change in pH is defined as a pH sensor sensitivity~\cite{bergveld2003thirty}. With this sensor, we achieved a sensitivity of the conventional Nernst limit of $\Delta V=$ \SI{59.2}{\milli\volt}$/$\si{pH} at a pH range of 4 to 9 and at the temperature of 30\textdegree{}C (Fig. \ref{fig:foobar}(a)).

\begin{figure}[!t]
    \includegraphics*[width=1\columnwidth]{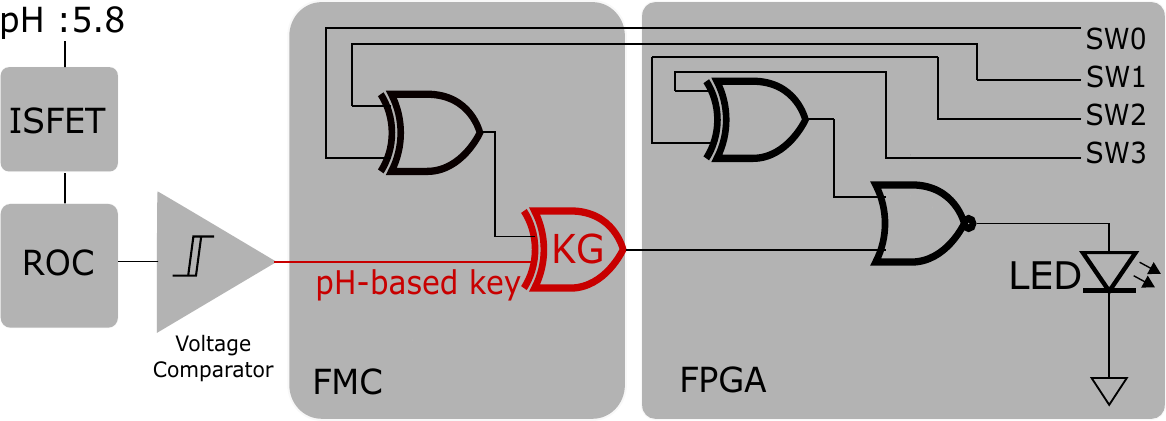}
  \caption{Locked circuit schematic and implementation on the FPGA board with an FMC connector.}\label{fig:logiclocking}
  \vspace{-6mm}
\end{figure}
The output voltage (the source voltage of the ISFET ($V_{S}$)) changes proportionally to the pH value (pH 4 to 9), as presented in Fig. \ref{fig:foobar}(a). The pH and temperature response of the ISFET over time are presented in Fig. \ref{fig:foobar}(b) for two pH (4 and 7) and temperature levels (30\textdegree{}C and 40\textdegree{}C). However, more sensitivity can be attained by enlarging the sensing gate of an extended gate ISFET~\cite{6479098} or double gate ISFET design (front and back oxide capacitance)~\cite{sanjay2021super}. To measure the potential developed in the ion-sensitive gate, an ISFET together with a reference electrode \ch{(Ag/AgCl)} must be immersed into the solution. For the solution, a phosphate buffer is prepared to adjust the pH value well between the pH range of 5.8 to 8.0, the example buffer that we used during the measurements is pH 5.8 with a voltage of \SI{2.07}{\volt} (Fig. \ref{fig:foobar}(a)). The reason to stay in the pH range of 5.8 to 8.0 is that many macromolecular properties are pH-dependent and stable within this range~\cite{talley2010ph}, thus enabling stable bio-based keys in the next stage. Table \ref{tab:rocchar} summarizes the specifications of the designed readout circuit for \si{pH} $= 5.8$. A microfluidic system can be used to control the flow of the pH solution to drive more logic gates.

\begin{figure}[!t]
   \centering
   \includegraphics*[width=1\columnwidth]{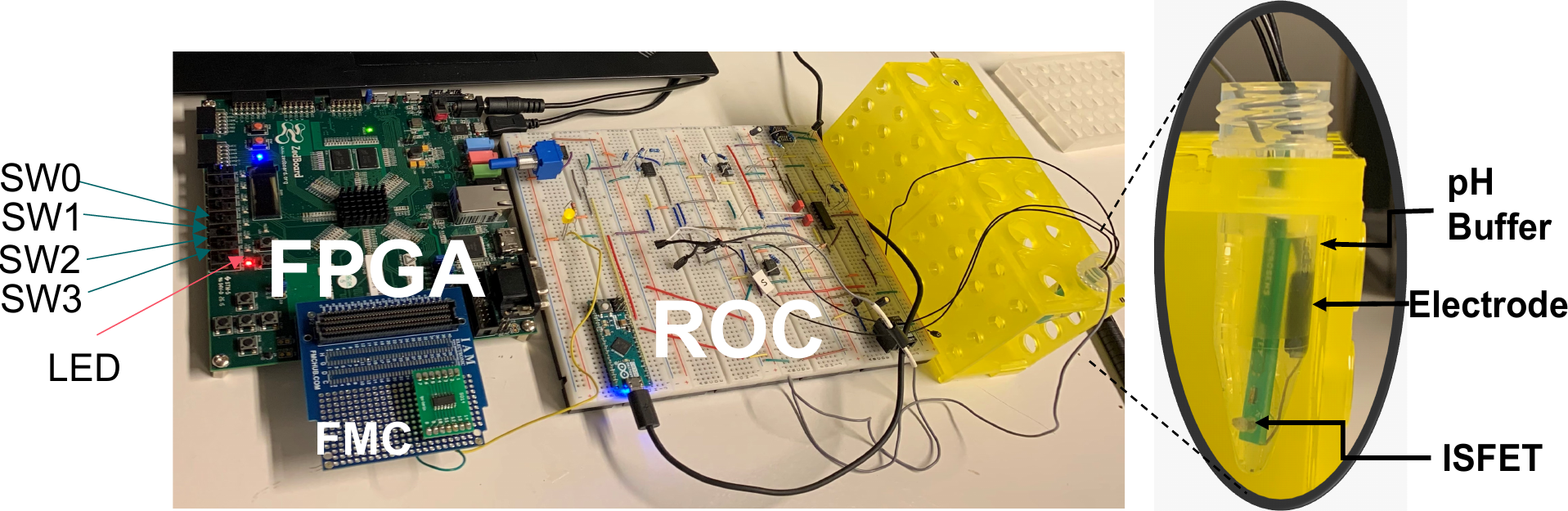}
   \caption{Demonstration of logic locking with pHGen.}
   \label{fig:prototype}
   \vspace{-0.3cm}
\end{figure}
To demonstrate the integration of pHGen with logic locking, a locked circuit was implemented by inserting a key gate (KG) in a selected gate-level netlist, as illustrated in Fig. \ref{fig:logiclocking}. The pH-based key for the key gate is provided by a pH buffer solution (\si{pH} $= 5.8$). The gate of the ISFET was immersed into the buffer solution to read ion activity in the form of a voltage shift. Thus, the CVCC circuit reads the amount of voltage drift and buffers it to isolate the driver from the load. A voltage comparator constantly compares the voltage with the known pH voltage value. Because the IP owner designs the circuit and knows the correct voltage change from the pH-based key. To have a functional circuit, the output voltage obtained from the pH solution needs to be equal to the same designed voltage value. This voltage sets the value of the input pin of the key gate (XOR KG) via an FPGA Mezzanine Card (FMC) connector. The locked circuit was implemented on a ZedBoard Zynq-7000 FPGA board connected to an FMC. Four switches (SW 0-3) are set as input signals to two XOR gates and combined with a NOR gate to implement a pHGen logic locked circuit. An LED is connected to the output pin of a NOR gate to indicate the output signal.
The FMC connector is used to bring modularity to change the I/O configuration of the design without FPGA intervention. Hence, a quad 2-input XOR gate IC is placed on the FMC connector. One of the XOR gates is used to establish a connection between the output voltage of the readout circuit and the input pin of the key gate. The FPGA-based test setup is shown in Fig. \ref{fig:prototype}. The pH-driven key gate is added to the netlist, through the readout circuit, via the FMC connector. 
To demonstrate the key generation, the ISFET together with the reference electrode are immersed in the specific pH solution (secret key) to create a particular voltage shift and activate the XOR gate with the dedicated key value.
 The switches (SW 0-3) are considered as input signals and the results for all input patterns are shown in Table \ref{tab:result}. The obtained results state that the pH-key generation is feasible, CMOS-compatible, and can produce the same behavior as a conventional binary key in the form of voltage shift. Due to the robustness of the CVCC readout circuit, the ISFET operates in the linear region at a fixed drain-source voltage, independent of temperature and the power supply variation. Under these conditions, the pH solution's ion activity was monitored, and the output voltage was observed as a change in the ISFET threshold voltage. With the presented use case, we demonstrated the feasibility of using chemical information as a source to secure analog keys.

\begin{table}[!t]
\caption{\label{tab:rocchar} 	
Readout circuit characteristics at 24\textdegree{}C, \si{pH}: 5.8.}
\centering
\begin{tabular}{l|c}
\hline
                 & Typ  \\ \hline
Supply voltage ($V_{CC}$) & \SI{0.5}{\volt}  \\ \hline
Operating current ($I_{DS}$) & \SI{50}{\mu\ampere}   \\ \hline
Internal drain voltage ($V_{D}$) & \SI{0.7}{\volt}  \\ \hline
Internal source voltage ($V_{S}$) & \SI{0.2}{\volt} \\ \hline
Output voltage ($V_{OUT})$ & \SI{2.07}{\volt} \\ \hline
CVCC circuit power consumption & \SI{50}{\mu\watt} \\ \hline
\end{tabular}
\vspace{-2mm}
\end{table}

\begin{table}[!t]
\caption{\label{tab:result} Results of the demonstrated locked circuit.}
    \centering
    \begin{tabular}{c|c|c}
    \hline
        \vtop{\hbox{\strut Input patterns}} & \vtop{\hbox{\strut Output (\si{pH} $= 5.8$,} \hbox{\strut Voltage = \SI{2.07}{\volt}, key =1)}} & \vtop{\hbox{\strut Output (\si{pH} $\neq 5.8$,} \hbox{\strut Voltage = \SI{0}{\volt}, key $\neq$ 1)}} \\ \hline
        0000 & 0 & 1 \\ \hline
        0001 & 1 & 0 \\ \hline
        0010 & 1 & 0 \\ \hline
        0011 & 0 & 1 \\ \hline
        0100 & 0 & 0 \\ \hline
        0101 & 0 & 0 \\ \hline
        0110 & 0 & 0 \\ \hline
        0111 & 0 & 0 \\ \hline
        1000 & 0 & 0 \\ \hline
        1001 & 0 & 0 \\ \hline
        1010 & 0 & 0 \\ \hline
        1011 & 0 & 0 \\ \hline
        1100 & 0 & 1 \\ \hline
        1101 & 1 & 0 \\ \hline
        1110 & 1 & 0 \\ \hline
        1111 & 0 & 1 \\ \hline
    \end{tabular}
    \vspace{-5mm}
\end{table}
\section{Next steps}
This novel approach of utilizing chemical information for securing ICs is expected to enable the application of next-in-line DNA-based nanoISFETs~\cite{pachauri2016biologically}~\cite{toumazou2014new} for personalized, secure systems. The work presented here establishes a basis for exploring the design of such chemically inspired secure ICs. In the next steps, we plan to extend the work to enhance the security aspects of logic locking in the following directions: (1) the design of multi-valued logic, (2) the implementation of massively parallel ISFET arrays to enable more key gates, and (3) the usage of nanoscale ISFETs to deploy super sensitive characteristics combined with high density and CMOS-compatible integration. (4) Finally, a microfluidic system will be developed to automatically control the flow of the solution.
\section{Conclusion}
In this work, we presented pHGen, a novel analog key generation approach that can be utilized as an alternative to conventional binary keys. Emerging technologies, such as the ISFETs, ameliorate the change towards the use of analog keys. The key is generated by shifting the intrinsic voltage threshold of the ISFET, which is observed based on the changes in the pH-soluble ion activities. To demonstrate the proposed mechanism, we integrated pHGen with logic locking---a prominent technique to protect hardware designs against manipulations and piracy. According to the results, pH-based analog keys have the ability to substitute conventional digital keys. Hence, pHGen provides a stepping stone to using a chemical-based key with ISFETs. 

\section*{Acknowledgement}
This work was partially funded by Deutsche Forschungsgemeinschaft (DFG – German Research Foundation) under the priority programme SPP 2253.

\bibliographystyle{IEEEtran}
\bibliography{bibliography}
\end{document}